\documentclass[prb,preprint,aps,showpacs]{revtex4}
\usepackage{graphicx}
\usepackage[english]{babel}
\usepackage{amscd}
\usepackage{tabularx}
\usepackage{latexsym}
\usepackage{amsmath}
\usepackage{amsfonts}
\usepackage{amssymb}
\usepackage[latin1]{inputenc}
\usepackage{epsfig}
\usepackage{times}
\usepackage[T1]{fontenc}
\usepackage{graphics}
\usepackage{verbatim}
\usepackage[absolute]{textpos}
\usepackage{wrapfig,times}
\usepackage{amsthm}
\usepackage{setspace}

\newcommand{\be}{\begin{equation}}
\newcommand{\ee}{ \end{equation}}
\newcommand{\ben}{\begin{eqnarray}}
\newcommand{\een}{\end{eqnarray}}
\newcommand{\sech}{\rm sech}

\begin{document}

\title{Thermoelectric Properties of $BiSbTe$ Alloy Nanofilms Produced by DC Sputtering: Experiments and Modeling}

\author{Andre A. Marinho$^{1,2}$, Neymar P. Costa$^{2}$, Luiz Felipe C. Pereira$^{2}$, Francisco A. Brito$^{1,3}$, Carlos Chesman$^{2}$}

\affiliation{$^{1}$ Departamento de F\'isica, Universidade Federal de Campina Grande, 58109-970 Campina Grande, PB, Brazil.
\\
$^{2}$ Departamento de F\'isica, Universidade Federal do Rio Grande do Norte, 59078-970 Natal, RN, Brazil.
\\
$^{3}$ Departamento de F\'isica, Universidade Federal da Para\'iba, Caixa Postal 5008, 58051-970 Jo\~ao Pessoa, PB, Brazil}

\date{\today}

\begin{abstract} 
Thermoelectricity refers to the conversion of thermal energy into electrical energy and vice-versa, which relies on three main effects: Seebeck, Peltier and Thomson, all of which are manifestations of heat and electricity flow. 
In this work we investigate the deposition of nanometric films and the effect of a thermal treatment on their thermoelectric properties. 
The films are based on $BiSbTe$ ternary alloys, obtained by deposition on a substrate using the DC sputtering technique. 
We produced sputtering targets with repurposed materials from commercial thermoelectric modules. 
In this way, we explore an environmentally responsible destination for discarded devices, with {\it in situ} preparation and manufacture of film-based thermoelectric modules.
Film samples show an improvement trend in thermoelectric efficiency as the annealing temperature is increased in the range 423 -- 623 K. 
The experimental data regarding thermal conductivity, electrical resistivity (or electrical conductivity), and the Seebeck coefficient were analyzed with the theory of $q$-deformed algebra.
Applying a $q$-deformation to our system we can model the effect of the annealing temperature on the thermal and electrical conductivities, as well as the Seebeck coefficient, and argue that the $q$-factor must be related to structural properties of the films.
We believe that our work could pave the way for future developments in the modeling of experimental measurements via the formalism of $q$-deformation algebra.
\end{abstract}


\maketitle


\section{Introduction}
\label{int}

In general, thermoelectricity refers to the conversion of thermal energy into electrical energy and vice-versa. 
The conversion relies on three transport effects: the Seebeck effect, the Peltier effect and the Thomson effect, all of which are manifestations of heat and electricity flow. 
Although any flow constitutes an irreversible process, the thermoelectric effects are thermodynamically reversible. 
In fact, those effects belong to an interesting class of physical phenomena arising from the junction of two or more irreversible processes. 
Even though the phenomenological theory of such processes cannot be rigorously described by conventional equilibrium thermodynamics, it can be described by a theory of reversible processes taking into account simultaneous changes of entropy due to reversible and irreversible energy fluxes \cite{groo}.

The performance of a thermoelectric material is characterized by a dimensionless parameter known as the figure of merit, $ZT$. 
This parameter brings together electrical and thermal characteristics in the form
\begin{equation}
\label{eq:zt}
ZT = \frac{\alpha^2}{\rho \kappa} \, T, 
\end{equation}
where $\alpha$ is the Seebeck coefficient, $\rho$ is the electrical resistivity, $\kappa$ is the thermal conductivity, and $T$ is the absolute temperature.
The combination $\alpha^2/\rho$ is known as the power factor, which is also an indicator of thermoelectric efficiency.
High performance thermoelectric materials, suitable for thermoelectric generators or heat pumping systems, present $ZT \ge 1$ at room temperature.

In the 1960's there was high demand for autonomous sources of electricity due to space exploration, advances in medical physics, and the exploitation of terrestrial resources in inaccessible locations \cite{row}. 
Thermoelectric generators are ideal for such applications, where reliability and absence of moving parts (enabling quiet operation) are advantages that compensate their relative high cost and low efficiency. 
The advantages of thermoelectric generators are notorious when compared to thermo-mechanical converters due to their simplicity and robustness \cite{coo}. 

During the last decade new materials such as skutterudites \cite{huan,kim}, and systems such as super networks \cite{har,ven} and nanostructured films \cite{har1,sha}, were developed to reach high $ZT$ values due to a reduction of their lattice thermal conductivity. 
However, these materials are expensive and require complex processing and manufacturing technologies. Furthermore, some of them are optimized for high temperature applications. 
To date, Bismuth-Antimony-Tellurium ($BiSbTe$) alloys remain one of the most suitable options for room temperature applications.
However, traditional thermoelectric systems employ these alloys in their bulk form, which is not appropriate for highly integrated systems due to mechanical limitations in microscale devices \cite{gro}. 

There is a large number of applications for thermoelectric materials in the form of films. 
For example, films can play the role of thermally active substrates which can be applied in the study of magnetocaloric and magnetogalvanometric effects \cite{gro}.
It would also be advantageous if thermoelectric modules could be made shorter while maintaining the dimensions of their cross section for commercial products. 
Therefore, there is a need to develop high quality thermoelectric thin films, with a figure of merit comparable to bulk materials, and compatible with microfabrication processes. 
The application of thin films in microsystems also presents several challenges, such as problems in the adhesion of the film to the substrate and the contact resistance between layers.
However, those issues did not prevent the manufacturing and use of thermoelectric micro-modules.

It is generally accepted that thermoelectric materials produced by a sintering process present better mechanical properties when compared to those produced by melt growth methods. 
Materials with smaller grain sizes present more inelastic phonon scattering, leading to a decrease in lattice thermal conductivity and consequently a higher $ZT$.
Several systems have already been successfully produced by sintering such as $SiGe$ \cite{vin}, $PbTe$ \cite{fan} and $Bi_2Te_3$, mostly following a procedure described by Cope and Penn \cite{cop}. 
Recently, polycrystalline or grainy composite materials have been shown to be superior to single crystals, and 
some of the best materials appear to be nanostructured bulk materials, in which the lattice thermal conductivity is reduced without affecting its electronic properties \cite{Felix2018}.

In this work we investigate the deposition of nanometric films and the effect of a thermal treatment on their thermoelectric properties. 
The films are based on $BiSbTe$ ternary alloys, obtained by deposition on a substrate using the DC sputtering technique. 
We produced targets for the sputtering system with repurposed materials from commercial thermoelectric modules. 
In this way, we explore an environmentally responsible destination for discarded modules, with {\it in situ} preparation and manufacture of film-based thermoelectric modules.
Film samples show an improvement trend in thermoelectric efficiency as the annealing temperature is increased in the range 423--623 K. 
The experimental data regarding thermal conductivity, electrical resistivity (or electrical conductivity), and the Seebeck coefficient were analyzed with the theory of $q$-deformed algebra.
Applying a $q$-deformation to our system we can model the effect of the annealing temperature on the thermal and electrical conductivities, as well as the Seebeck coefficient, and argue that the $q$-factor must be related to structural properties of the films.

The paper is organized as follows. 
In section \ref{etp} we present the experimental techniques used to produce thermoelectric thin films. 
In section \ref{res} we present and discuss the experimental results. 
In section \ref{qdqa} we introduce the $q$-deformed algebra and model the experimental data.
Finally, in section \ref{con} we present our concluding remarks.

\section{Experimental Techniques and Procedures}
\label{etp}

Semiconducting alloy films can be produced by thermal evaporation of elemental sources in vacuum. 
However, it is often difficult to control the deposition rate of different compounds during the deposition process, especially if they have a wide difference in melting points. 
Thus, it is more common to evaporate from a material which already has a composition close to the final product. 
Flash Evaporation is a technique which ensures that the deposited film has almost the same chemical 
composition as the source. 
Takashiri showed that a $Bi_{0.4}Sb_{1.6}Te_{3}$ film produced by this technique may have a power factor of $3.5$ mW m$^{-1}$ K$^{-2}$, which is comparable to $4.0$ mW m$^{-1}$ K$^{-2}$ presented by the bulk material in a specific direction \cite{tak}. 
Meanwhile, cathodic evaporation, or sputtering, is extensively used in the semiconductor industry for the deposition of thin films of various materials. 

We produced p-type $BiSbTe$ targets by reusing the semiconducting material from thermoelectric modules, via grinding, sieving, compaction and cold sintering.
Ater that we prepared a series of p-type semiconducting films in glass substrates by DC magneton sputtering from the $BiSbTe$ targets.
The deposition rate of the films was controlled by adjusting the power of the DC source as well as the $Ar$ pressure.
For each film, the deposition rate was determined by X-ray diffraction (XRD) analysis at low angles (2 to 5 degrees) at consistent time intervals.
The morphological characterization of the films was performed via XRD and X-ray fluorescence.

Unless the film substrate is heated, chemical reactions may not occur and annealing should be employed. 
Annealing is also necessary in order to achieve uniformity and lower the density of structural defects in the films.
One of our objectives was to study the effect of heat treatments after deposition on the thermoelectric properties of the films, as well as compare our results to what is available in the literature \cite{ger,vil,bou}.
Therefore, after deposition, $BiSbTe$ films were thermally treated by annealing in $Ar$ atmosphere for $60$ minutes \cite{tak}.
A limiting temperature of $623$ K was established, since higher temperatures would cause evaporation of more volatile components in the film structure, such as $Te$ itself. 
Another factor limiting the annealing temperature is the adhesion of the film to the substrate, since higher temperatures can cause detachment from the substrate.

Meanwhile, Seebeck coefficient, thermal conductivity and electrical resistivity were measured with a Physical Property Measurement System (PPMS) from Quantum Design in the termal transport option mode (TTO mode). 
In this operation mode the system performs an electrical resistivity measurement and then promotes a heat pulse generated by a heater coupled at one end of the sample. 
The heat flow is directed to the other end which is connected to the thermal reservoir. 
The pulse duration and intensity is determined by a software which extracts the value of the thermal constants by extrapolation and least squares adjustments. 
The heat pulse originates temperature gradient, and the thermal conductance is obtained from the temperature measurement in the thermometers.
Simultaneously, the Seebeck voltage is measured by the same thermometer terminals, in one cycle of the heat pulse.
The properties are sensitive to small temperature variations, which is controlled with a $0.01\%$ precision in each cycle.
All measurements were performed at room temperature. 
Finally, the figure of merit is calculated from the measured properties with Eq. (\ref{eq:zt}).

\section{Results and discussion}
\label{res}

\begin{figure}[htb]
\centerline{
\includegraphics[width=0.75\linewidth]{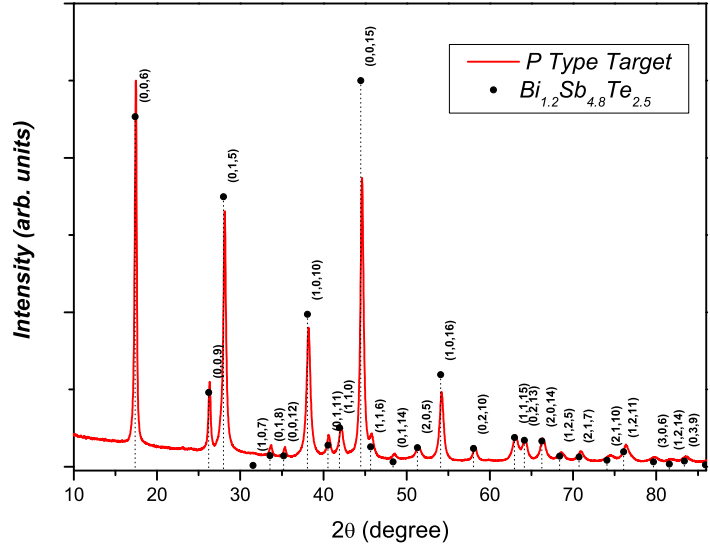}
}
\caption{XRD pattern of targets and peaks relating to the phase of the $Bi_{2-x}Sb_xTe_3$ ternary alloy, letter $03-065-3674$ JCPDS. }
\label{fig01}
\end{figure}

Currently, $BiSbTe$ ternary alloys are the compounds which present the highest $ZT$ for thermoelectric applications at room temperature \cite{dug}. 
In general, these alloys present a $Bi_{2-x}Sb_xTe_3$ stoichiometry. 
Fig. \ref{fig01} shows a diffractogram of the targets produced according to our method.
The agreement between the peaks in the target (continuous line) and the dots indicate the successful growth of $Bi_{1.2}Sb_{4.8}Te_{2.5}$ films.

\begin{figure}[htb]
\centerline{
\includegraphics[width=0.75\linewidth]{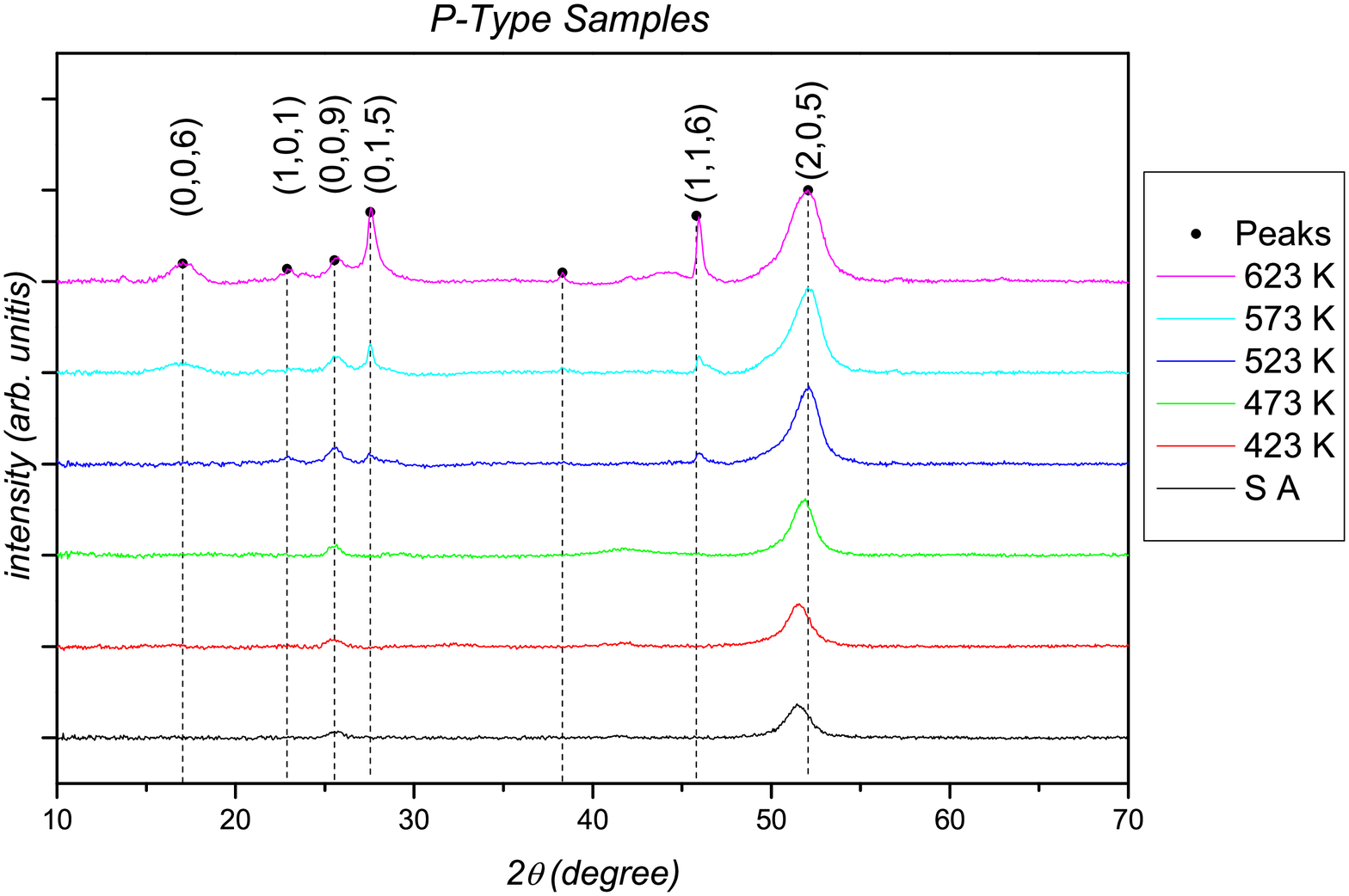}
}
\caption{Diffractograms of deposited thermoelectric film samples produced for each annealing temperature. SA indicates the non heat treated sample. The peaks are indexed according to the letter $03-065-3674$ JCPDS.}
\label{fig02}
\end{figure}

In Fig. \ref{fig02} we show the XRD pattern of p-type thermoelectric films deposited. 
The increasing number of peaks in the XRD spectra as the annealing temperature is increased from bottom to top is taken as evidence of the crystallization process due to heat treatment in those samples.
Therefore, we can assert that the annealing temperatures chosen are effective in increasing the crystallinity of the films.

Measurements of the physical properties for each one of the samples, obtained from the thermal transport module of the PPMS at room temperature, are presented in Tab. \ref{tab01}: thermal conductivity $\kappa$, Seebeck coefficient $\alpha$, electrical resistivity $\rho$ and figure of merit $ZT$. 
It is important to highlight that measuring $ZT$ with high precision is a real challenge and, in general, uncertainties can be as larger as 20\% \cite{gol1}. 
We observe a modest monotonic increase in $\kappa$ and a two orders of magnitude decrease in $\sigma$, which causes a two orders of magnitude increase in $ZT$ relative to the sample with no heat treatment (P104).
The dependence of $\alpha$ on the annealing temperature appears somewhat more complex.
We observe that, although the compound of the target is a ternary compound of p-type bismuth telluride, $\alpha$ presents negative values for the first samples, P104 and P105, without annealing and treated at $423$ K.
In principle, negative $\alpha$ values are compatible with n-type semiconductors. However, it has been reported in the literature that $n-p$ type transitions can occur in $BiSbTe$ ternary compounds \cite{mul}.
Considering samples subject to an annealing temperature of $473$ K and more, we observe a decrease in the Seebeck coefficient, which is an indicative that the structures formed in the deposition are polycrystalline nanocomposites, since for these structures, $\alpha$ falls continuously \cite{kim1}.

\begin{table}[htb!]
\centering
\begin{tabular}{|c|c|c|c|c|c|c|}
\hline
{Sample} & {Ann. (K)} & 
{Sample Temp.(K)} & 
{$\kappa ^{(a)}$} & 
{$\rho^{(b)}$} &
{$\alpha ^{(c)}$} &
{$ZT^{(d)}$}
\\
\hline
P104 & -- & 304.15 & 0.835 & 3.236$E-01$ & -122.60 & 1.57$E-05$ \\
\hline
P105 & 423 & 304.74 & 0.882 & 2.96$E-01$ & -11.38 & 1.42$E-07$ \\
\hline
P107 & 473 & 303.88 & 0.932 & 1.05$E-01$ & 409.58 & 5.64$E-04$ \\
\hline
P109 & 523 & 304.05 & 1.036 & 1.14$E-02$ & 254.76 & 1.68$E-03$ \\
\hline
P110 & 573 & 304.37 & 1.071 & 9.59$E-03$ & 316.58 & 2.90$E-03$ \\
\hline
P111 & 623 & 304.41 & 1.178 & 4.54$E-03$ & 145.46 & 1.03$E-03$ \\
\hline
\end{tabular}
\caption{{Results of thermoelectric measurements for samples: thermal conductivity $^{a}${$\left({W}{m^{-1} K^{-1}}\right)$}, electrical resistivity $^{b}$ ($\Omega \cdot m)$, Seebeck coefficient $^{c}${$(\mu V/K)$}, and figure of merit $^{d}$. 
}}\label{tab01}
\end{table}

The Seebeck coefficient and the electrical resistivity are combined to form the power factor $\alpha^2/\rho$.
Notice that a low electrical resistivity also implies an increase of the electronic contribution to the thermal conductivity $\kappa$, which could consequently decrease $ZT$.
A restructuring of the material due to the heat treatment can significantly alter the energy gap for $Bi_2 Te_3$ compounds which at $300$ K is around $5 \kappa_B T$ (where $\kappa_B$ is Boltzmann constant). 
This means that for typical samples, minority charge carriers cannot be totally neglected. 
Minority carriers have thermoelectric coefficients that are opposite in signal to those of the majority carriers, thus reducing the total Seebeck coefficient as well as contributing strongly to the thermal conductivity by the bipolar effect \cite{gol}.

\begin{figure}[htb]
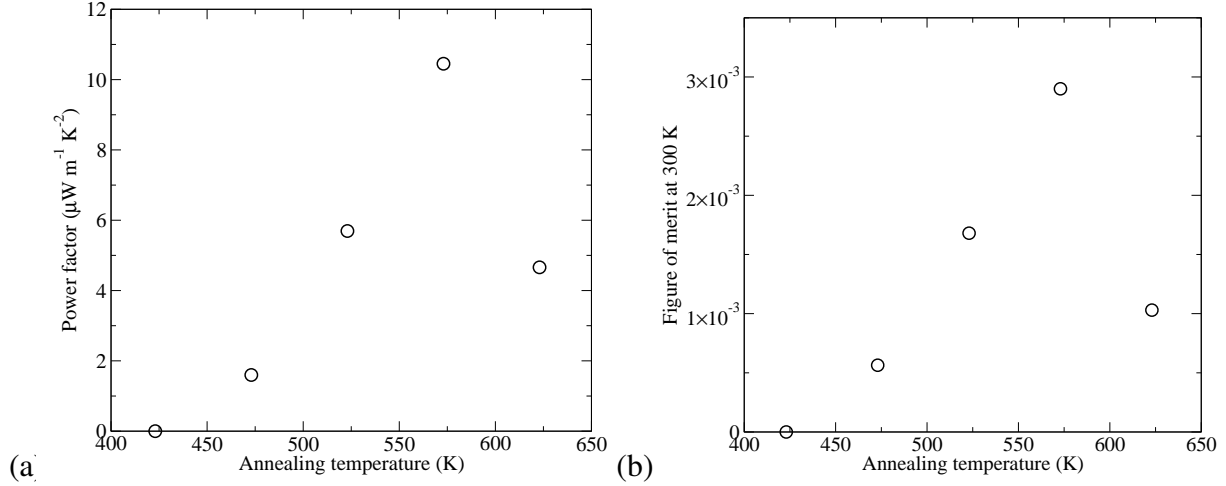

\centerline{
(a) \includegraphics[width=0.45\linewidth]{fig03a.eps}
(b) \includegraphics[width=0.45\linewidth]{fig03b.eps}
}\caption{(a) Power factor and (b) thermoelectric figure of merit, as a function of the annealing temperature in the range from 423 -- 623 K. Both quantities reach a maximum for an annealing temperature of 573 K.}
\label{fig03}
\end{figure}

The data in Fig. \ref{fig03}(a) and (b) shows an improvement in the power factor and $ZT$ up an annealing temperature of $573$ K. Therefore, this would be the ideal annealing temperature to enable the highest thermoelectric efficiency.
Also, the observed behavior is compatible with polycrystalline films, whose grain sizes increase as the annealing temperature is increased.
Indeed, Hicks and Dresselhaus \cite{hic} observed a reduction of the lattice thermal conductivity when nanostructures were used in the preparation of the materials, although their original objective was to improve the power factor through quantum confinement effects. 
In fact, it is more advantageous to have phonon scattering at the boundaries of nanometric structures, so that nanostructuring appears to affect the lattice thermal conductivity while not interfering with the properties of electronic transport.
The decrease in the power factor as we go to an annealing temperature of $623$ K could be due to undesired thermal degrading effect in the samples.

In what follows, we propose a model to the experimental data obtained for electrical and thermal conductivities, as well as the Seebeck coefficient itself, through a $q$-deformed algebra applied to a Debye solid.
Our model indicates a direct relation between the $q$-deformation factor and the annealing temperature of the samples. 

\section{$q$-deformation model}
\label{qdqa}

In the literature, several generalized statistical models have been intensively investigated \cite{gen,gre,pol,bie,mac,rpv,chai,fuc,lav,gav2,hat,aba,hoy,gav,tsa,arp,sam}. 
In our model, we begin by defining the quantity know as the {\it basic number} \cite{erns}, $[n]$, in terms of the creation and annihilation operators ($a^{\dagger}$ and $a$), as well as the number operator $N$
\be
\label{eq:BN}
[n] \equiv a^{\dagger}a=\frac {q^{{2N}}-q^{-{2N}}} {q^{{2}}-q^{-{2}}},
\ee
where $q$ is a deformation parameter which modifies the physical properties of the system.
Within the $q$-deformed algebra, the quantum harmonic oscillator is now described by the Heisenberg equation in the form
\be
aa^{\dagger} - q^{-2}a^{\dagger}a= q^{{2N}},
\ee
where $a$, $a^{\dagger}$ and $N$ satisfy the usual commutation relations
\be 
[a,a] = [a^{\dagger},a^{\dagger}] = 0, \; [N,a^{\dagger}]= a^{\dagger}, \; [N,a] = -a.
\ee
Consequently, we have
\be
aa^{\dagger}=[1+n].
\ee

Now we have the following Hamiltonian and the respective energy eigenvalues
\begin{equation}
 \label{eq1}{\cal H} = \frac{\hbar\omega}{2}\left(a^{\dagger}a+aa^{\dagger}\right),\qquad\quad 
E_n = \frac{\hbar\omega}{2}\Big([n]+[n+1]\Big).
\end{equation}
Taking the definition of the {\it basic number}  in Eq. (\ref{eq:BN}), and making $q=\exp(\gamma)$ \cite{flo}, we obtain 
\begin{equation} 
\label{eq5} E_{n_q} = \frac{\hbar\omega}{2}\left[\frac{\sinh\left(\left(n+\frac{1}{2}\right) {2\gamma}\right)}{\sinh\left({\gamma}\right)}\right], \qquad  
[n] = \frac{\sinh\left({{2n\gamma}}\right)}{\sinh\left({2\gamma}\right)}. 
\end{equation} 

\subsection{Implementation of the $q$-deformation}
\label{iqd}

The Debye model describes the vibrational modes of a solid in terms of a continuous frequency spectrum up to a cut-off frequency, $\omega_D$, such that the total number of normal modes equals that of the solid being described \cite{patt,reif,hua,kit,zim}.
According to the mathematical development above, we can employ the parameter $q$, and write the specific heat at a given temperature as 
\begin{equation} 
\label{eq:cv}
c_{V_q}(T) = \frac{9 \kappa_B }{(\theta_{D_q}/{T})^3} \int_{0}^{\theta_{D_q}/{T}} \frac{x^4\exp(x)} {[\exp(x)-1]^2}dx
\end{equation}
where $\kappa_{B}$ is the Boltzmann constant, $\theta_{D_q}$ is the $q$-deformed Debye temperature defined as
\begin{equation}
\theta_{D_q} = \theta_D \; \frac{\sinh\left[2\ln(q)\right]} {2\sinh\left[\ln(q)\right]}=\theta_{D}\cosh[\ln(q)],
\end{equation} 
where $\cosh(\phi)$ is the hyperbolic cosine function.

Integrating Eq.~(\ref{eq:cv}) by parts, we can verify that the low-temperature specific heat in a $q$-deformed Debye solid is 
proportional to $T^3$, as in the usual Debye solid.
For low-temperatures, $T \ll \theta_{D_q}$, such that $\theta_{D_q}/{T} \gg 1$, and we can perform the integration to write the 
$q$-deformed specific heat as 
\begin{eqnarray}
\label{eq.24} c_{V_q} = \frac{12\pi^4\kappa_B}{5}\left(\frac{T}{\theta_{D_q}}\right)^3.
\end{eqnarray}
This predicted low-temperature behavior is due to phonon excitations, and is in agreement with experimental observations. 

We can relate the $q$-deformed specific heat to its non-deformed counterpart by 
\be
c_{V_q}=c_{V} \sech^3 [ \ln(q) ],
\ee
where $\sech(\phi)$ is the hyperbolic secant function.
Meanwhile the $q$-deformed and non-deformed conductivities are related by \cite{zim},
\be 
\label{eq.25}
\kappa_q= \kappa \, \frac{c_{V_q}}{c_{V}}, \;  \sigma_q=\sigma \, \frac{\kappa_{q}}{\kappa}.
\ee
Finally, we can write 
\begin{equation} 
\label{eq.26} 
\kappa_{q}=\kappa\,\sech^3[\ln(q)], \; \sigma_q=\sigma\,\sech^3[\ln(q)].
\end{equation}

Ordinarily, the Seebeck coefficient is given by the entropy to charge ratio.
Now, we can take one step further and obtain a $q$-deformed expression for the Seebeck coefficient, given by
\be
\alpha_q=\frac{S_q}{Q},
\ee
where $S_q$ is the $q$-deformed entropy and $Q$ is the charge, which is negative for electrons and positive for holes.
From thermodynamics we know that specific heat and entropy variation are related, therefore we can write for the $q$-deformed quantities
\be
c_{V_q} = T \left( \frac{\partial S_q}{\partial T} \right),
\ee
which can be integrated to give $S_q$.
Finally, from the definition of the $q$-deformed Seebeck coefficient we obtain
\be
\alpha_q = \frac{4 \pi^4 k_B}{5 Q} \left(\frac{T}{\theta_D}\right)^3 {\sech^3[\ln(q)]} + \frac{C_{\alpha}}{Q},
\ee
where $C_\alpha$ is a constant of integration. The previous expression can be condensed in the form 
\be \alpha_q = \left[\frac{c_{V_q}}{3}+C_{\alpha}\right]\frac{1}{Q}.\ee
or equivalently
\be 
\alpha_q = \alpha\;\sech^3[\ln(q)]+\frac{C_{\alpha}}{Q}. 
\label{eq:alphaq}
\ee
In what follows, the above expressions will be fitted to the experimental data.

\subsection{Direct comparison between measurements and $q$-deformed model}
\label{cbr}

Let us now try to reproduce the experimental data for the conductivities and the Seebeck coefficient with the respective values calculated applying the $q$-deformation model.
The application of the $q$-deformed algebra to solids had already been demonstrated theoretically, where we suggested a possible experimental interpretation of the $q$-parameter, relating it to some type of perturbation or disorder in the system \cite{bri}.
Also, it is understood that the annealing process reduces the disorder in the samples, improving its crystallinity.
Therefore, we relate the $q$ parameter to the annealing temperature through an equation of correspondence, and write 
\be
\label{eq.28} 
q = \left(\frac{|T_i -T_A|}{300 K}\right)^{A_i},
\ee
where $T_A$ is the annealing temperature, $T_i$ and $A_i$ are adjustable parameters.
Now we can rewrite Eqs. (\ref{eq.26}) and (\ref{eq:alphaq}) as 
\be
\kappa_{q}=\kappa\sech^3\left[A_1\ln\left(\frac{|T_1-T_A|}{300 K}\right)\right]
\label{eq:kq}
\ee
\be
\sigma_q=\sigma\sech^3\left[A_2\ln\left(\frac{|T_2-T_A|}{300 K}\right)\right]
\label{eq:sq}
\ee
\be
\alpha_q = \alpha\sech^3\left[A_3\ln\left(\frac{|T_3-T_A|}{300 K}\right)\right]+\frac{C_{\alpha}}{Q}.
\label{eq:aq}
\ee
Naturally, when $| T_i-T_A |=300$ K we recover the undeformed case since for $q=1$ we have $\kappa_q=\kappa$, $\sigma_q=\sigma$, and $\alpha_q = \alpha$.

\begin{figure}
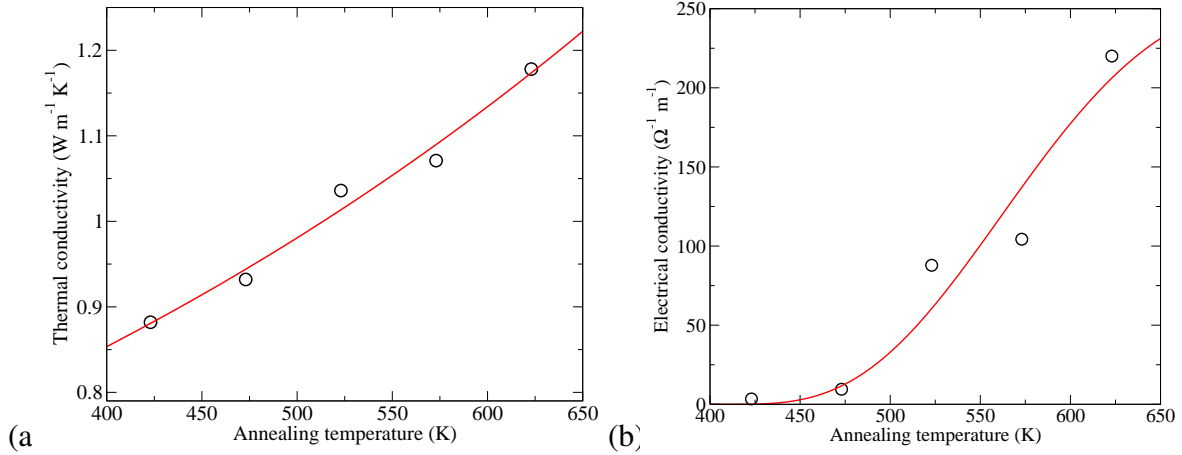

\centering
(a)\includegraphics[width=0.45\linewidth]{fig04a.eps}
(b)\includegraphics[width=0.43\linewidth]{fig04b.eps}
\caption{(a) $\kappa$ and (b) $\sigma$ as a function of annealing temperature. Data points represent the experimental measurements, while the continuous line is the least-squares fit to the $q$-deformation model.}
\label{fig04}
\end{figure}

In Fig. \ref{fig04}(a) and (b), the circles represent the experimental data, while the continuous line is given by Eqs. (\ref{eq:kq}) and (\ref{eq:sq}) with the adjusted parameters.
A least-squares fitting to the experimental data gives $A_1=-0.69$, $T_1=1481$ K, $A_2=-1.08$, and $T_2=410$ K.
It is noticeable that both sets of data are properly described by our model.
Even though the description of $\sigma$ is somewhat qualitative, the data for $\kappa$ is reproduced with a much better precision.
Nonetheless, the continuous line describing the behavior of $\sigma$ predicts a maximum at a higher annealing temperature, consistent with a degradation of the sample.

\begin{figure}
\centering
\includegraphics[width=0.7\linewidth]{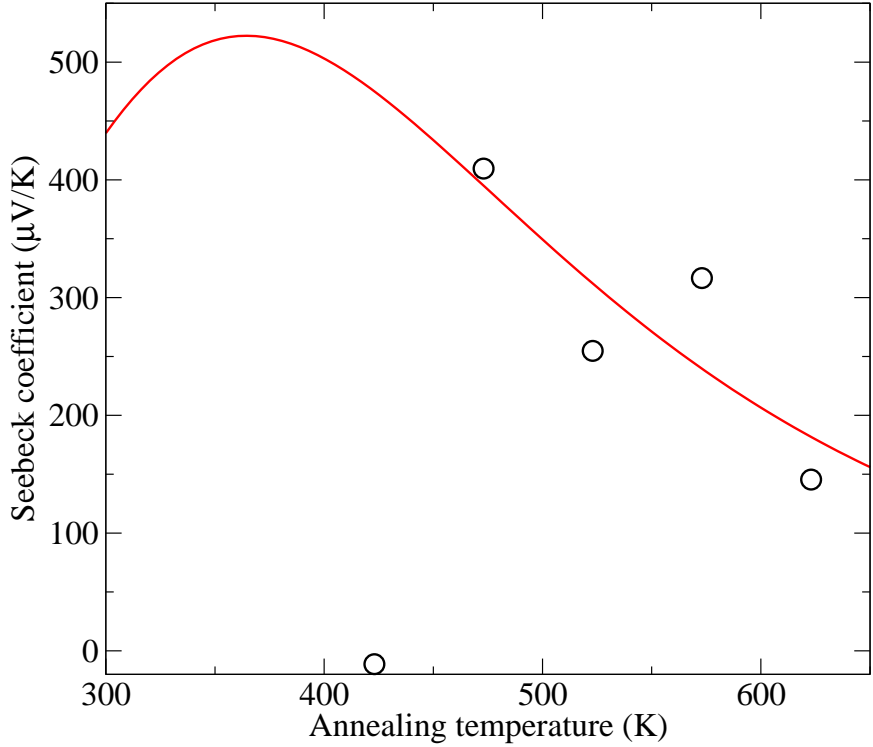}
\caption{Seebeck coefficient, $\alpha$ as a function of annealing temperature. Data points represent the experimental measurements, while the continuous line is the least-squares fit to the $q$-deformation model.}
\label{fig05}
\end{figure}

In the case of the Seebeck coefficient, shown in Fig. \ref{fig05}, the $q$-deformed expression given in Eq. (\ref{eq:aq}), also reproduces qualitatively the experimental data, with parameters $A_3=1.40$, $T_3=64.4$ K, and $C_{\alpha}/Q = -11.6$ $\mu$V/K.
Furthermore, the model predicts a maximum for an annealing temperature between 300 and 400 K. 
This maximum is consistent with the values reported in Tab. \ref{tab01}, where $\alpha$ changes sign as the annealing temperature is increased.

Finally, it is understood that the annealing process increases the crystallinity of the samples, in accordance with our XRD data.
Therefore, Eq. (\ref{eq.28}), relating the parameter $q$ to the annealing temperature, is in fact a relation between this somewhat abstract deformation parameter and the degree of crystallinity of the films.
In fact, this is the first time that a relationship between the $q$-deformation parameter and the structural properties of a real physical system has been found.
We believe that our work could pave the way for future developments in the modeling of experimental measurements via the formalism of $q$-deformation algebra.

\section{Conclusions}
\label{con}
The $BiSbTe$ films investigated in this work, prepared by the DC sputtering deposition technique, presented results in accordance with current  literature \cite{bou1,huan1}.
However, the great differential is that this was achieved using non-commercial targets, with repurposed material from comercial thermoelectric modules, explore an environmentally responsible destination for discarded devices.
We believe it is possible to improve the thermoelectric parameters of the films and, perhaps, produce films for thermally active substrates. 
For example, our thermoelectric films could be used to study magnetic properties of materials which show signs of galvanometric magneto effects and spin-related thermoelectric effects, since they would allow a localized control of temperature and of the thermal gradient.
The application of the $q$-deformed algebra to solids had been demonstrated previously, where we suggested a connection between the $q$-parameter and some type of perturbation or disorder in the system \cite{bri}.
Since the annealing process reduces the disorder in the samples, improving its crystallinity, we propose to relate the $q$ parameter to the annealing temperature. 
Our results show that we have a theoretical tool that can describe, at least qualitatively, experimental measurements such as electrical and thermal conductivities, as well as the Seebeck coefficient. 
This further motivates us to look for applications elsewhere, where the $q$-deformation algebra could be seen as  a tool to reduce costs and production time of thin films.


\acknowledgments

We would like to thank CAPES, CNPq (Grants 309961/2017-3, 436859/2018-1, 312104/2018-9) and PNPD/PROCAD-CAPES, for financial support.


\end{document}